\documentclass[amsmath,twocolumn,superscriptaddress]{revtex4-1}
\usepackage{epsfig}
\usepackage{amssymb}
\usepackage{amsmath}
\usepackage{amsfonts}
\usepackage{braket}
\usepackage[T1]{fontenc}
\usepackage{bm}
\usepackage{graphicx}
\usepackage{color}
\usepackage{verbatim}
\usepackage{cancel}
\usepackage{soul}

\begin{document}
\title{The Lyapunov Spectrum of Quantum Thermalisation}

\author{A. Hallam}
\affiliation{London Centre for Nanotechnology, University College London, Gordon St., London, WC1H 0AH, United Kingdom}

\author{J. Morley}
\affiliation{London Centre for Nanotechnology, University College London, Gordon St., London, WC1H 0AH, United Kingdom}

\author{A.~G. Green}
\affiliation{London Centre for Nanotechnology, University College London, Gordon St., London, WC1H 0AH, United Kingdom}

\date{\today}
\begin{abstract}
The eigenstate thermalisation hypothesis resolves the paradox of emergent thermal or classical behaviour in a closed quantum system by focussing upon local observations. This permits the remainder of the system to act as a bath, thermalisation arising due to a process of de-phasing that gradually reveals the thermal nature of local observables measured in an eigenstate. This is very different from thermalisation in closed classical systems, which is driven by dynamical chaos. We show how quantum thermalisation in closed systems can be recast in a way that is directly related to classical thermalisation.  Local observables can be accurately captured by projecting states onto a suitable variational manifold. Evolving on this manifold using the time-dependent variational principle projects the quantum dynamics onto a (semi-)classical Hamiltonian dynamics. Thermalisation in this setting is driven by dynamical chaos. We carry out this procedure for an infinite spin chain in two ways --- using the matrix product state ansatz for the wavefunction and for the thermofield double purification of the density matrix --- and extract the full Lyapunov spectrum of the resulting chaotic dynamics. This provides an alternative perspective upon eigenstate thermalisation, pre-thermalisation and integrability. 
\end{abstract}
\maketitle
\tableofcontents
 
\section{Introduction} 
The extra information required to specify a pure quantum state  compared to that required for a  classical or thermal state underpins many of the apparent paradoxes of quantum mechanics\cite{gogolin2016equilibration}. These may be profoundly philosophical, such as when attempting to apply quantum mechanics to the whole universe, {\it e.g.} the black hole information paradox, and the very long scale entanglement implied by the origin of microwave background anisotropy in zero-point fluctuations\cite{Susskind:2013rc}. Whilst there are fewer philosophical difficulties in the description of finite quantum systems, there are practical consequences. We focus upon one, namely the thermalisation of local observables in the quantum evolution of a closed system. 

Accurate numerical description of a quantum system evolving from a weakly entangled initial state requires an exponentially growing number of parameters. The eigenstate thermalisation hypothesis implies that, beyond a certain point in time, an accurate representation of this dynamics should require a reducing number of parameters. The eigenstate thermalisation hypothesis\cite{Deutsch:1991rx,Srednicki:1994qp,Rigol:2008zf} has made great strides in demonstrating how thermal correlations present in local observations of eigenstates are revealed through a process of de-phasing due to entanglement with regions of the system not directly under observation. The ultimate consequence is that late-time, local observations are characterised by just the energy density. 
Evidently, the reduction of parameters required to capture the late-time dynamics\cite{Page:1993dq,Page:1993tg}  is related to the emergence of classicality in local observations of the closed quantum system\footnote{The increase and then reduction of parameters required to accurately describe a thermalising quantum system is akin to the Page curve\cite{Page:1993dq,Page:1993tg} for the entanglement entropy of partitions of a system. The Page curve often appears in the context of the black hole information paradox and indeed, its appearance here is for very similar reasons; the difference being that the horizon for observations is imposed by hand in our case and does not evaporate.}. Here we demonstrate a new way to analyse quantum thermalisation that 
extends the connection between classical and quantum thermalisation. The central idea is to project the quantum dynamics onto an effective semi-classical, Hamiltonian dynamics on variational manifolds\cite{HaegemanTDVP,leviatan2017quantum}. Thermalisation in these classical systems occurs {\it via} dynamical chaos\cite{CalculatingLyapunov,1980Mecc...15....9B,RevModPhys.57.617}, which we characterise by extracting the full Lyapunov spectrum. 

We apply this reasoning to a translationally invariant spin chain, a system over which we have analytical and numerical control using matrix product state (MPS) representation of the wavefunction\cite{Orus:2014zl} and the thermofield double purification of the density matrix. In both cases, we follow the dynamics using the time-dependent variational principle.Amongst our key results, in the case of wavefunction MPS we find a zero-parameter fit between the Lyapunov spectrum and the time-dependence of entanglement. In the thermofield MPS near the centre of the spectrum, we recover a semi-circular distribution of Lyapunov exponents for thermalising systems, as found previously in the case of gravitational systems\cite{2016JHEP...02..091G,2018PhRvE..97b2224H}, and a Gaussian distribution for integrable systems.

By bringing the study of many-body quantum chaos into contact with that of classical chaos, our approach opens up the full range of techniques available in the latter. For example, it allows the potential to examine how the classical KAM theorem for deformations from integrable behaviour may manifest in quantum systems. It also suggests natural possibilities for efficient descriptions of late-time dynamics. This complementary perspective brings the study of quantum chaos full circle, recapitulating the characterisation of few particle quantum chaos through its projection to classically chaotic systems.

In Section \ref{Sec:ClassicalQuantumThermo}, we begin by reviewing the role of classical dynamical chaos in enabling the ergodicity and thermalisation of classical closed system. This introduces some of the ideas, methods, and nomenclature of classical chaos that we will later apply to projected quantum dynamics. We then turn to quantum dynamics and give  brief expositions of eigenstate thermalisation, the important role played by quantum chaos and the conventional characterisation of the latter through the eigen spectrum. In Section \ref{Sec:ProjectingQuantumDynamics}, we discuss the projection of pure quantum dynamics to a variational manifold, the conditions under which this captures the dynamics of a restricted set of observations, and how this projected dynamics reduces to an effective (semi-)classical dynamics. Section \ref{Sec:QuantumLyapunov} gives some of the technical detail (expanded upon in appendices) of how to extract the Lyapunov spectrum of projected quantum dynamics. Section \ref{Results} summarises our numerical results and the relationship of the Lyapunov spectrum to other measures of quantum chaos. Finally, we discuss the broader implications of our results.

\section{Classical and Quantum Thermalization} 
\label{Sec:ClassicalQuantumThermo}

\subsection{Classical Thermalization}
{\it Ergodicity and Chaos:}
Thermalisation in closed classical systems occurs due to dynamical chaos. Every dynamical mode of the system is chaotic, revealed on timescales given by the inverse of its corresponding Lyapunov exponent. On the longest timescales, evolution leads to ergodic exploration of states in phase space with a given energy (or other conserved quantities). This picture of classical thermalisation requires an explicit ensemble or time-averaging to obtain thermal averages, a point that we will return to later. On these timescales, only conserved quantities can be used to distinguish states of the system - small differences in states with the same values of conserved quantities are eventually randomised by the chaotic evolution and so cannot be used to characterise the state.  

On shorter timescales, dynamical modes can be divided into two classes; those that have revealed their chaotic behaviour, and those that have not. We will refer to these as chaotic and residual regular modes (a classification that is determined by a choice of timescale). Residual regular modes can be used to discriminate between states of the system on a given timescale. The chaotic modes effectively form part of a thermal bath, and time or ensemble averaging will draw uniformly from the possible amplitudes of deformations in chaotic directions. On increasing timescales, the number of residual regular modes decreases monotonically, until ultimately only conserved quantities remain as distinguishing features of classical states.
It is important to note that this is not simply a matter of averaging out high-frequency modes. A high frequency mode with high quality factor can  distinguish states on timescales longer than its period. Of course, it is plausible that frequencies and decay rates may be linked in some cases, but this is not necessarily so.

{\it The Lyapunov Spectrum:}
After setting the scene in this way, we now give an overview of how the Lyapunov spectrum is extracted for a classical dynamical system. There are many excellent reviews of this subject \cite{CalculatingLyapunov,1980Mecc...15....9B,RevModPhys.57.617}. We confine ourselves to a brief outline in order to contextualise our analysis of the quantum system.
Consider a dynamical system whose parameters are contained in a vector ${\bf X}(t)$ that evolves according to 
\begin{equation}
\partial_t {\bf X}(t) = {\bf F}({\bf X}(t)).
\label{eq:Classical EoM}
\end{equation}
The Lyapunov spectrum is found by considering the evolution of the displacement between neighbouring trajectories ${\bf X}(t)$ and ${\bf X}(t)+d{\bf X}(t)$, where $d{\bf X}$ is initially small. Expanding Eq.(\ref{eq:Classical EoM}) to leading order, we obtain the following equation for the evolution of the displacement between trajectories:
\begin{equation}
\partial_t d X_i 
=\partial_j F_i({\bf X}) dX_j.
\label{eq:Classical lEoM}
\end{equation}
The formal solution of this equation,
$d{\bf X}(t)=T( \exp [ \int_0^t {\bm \partial} {\bf F}(t') dt'] )d{\bf X}(0)$,
shows that instantaneously, $d{\bf X}$ grows exponentially and decreases exponentially with $t$ in the eigen-directions of $\partial_i F_j$. The exponents for this growth and decay are the instantaneous Lyapunov exponents and the Lyapunov exponents are the time average of them along a trajectory. These equations are manipulated in various ways to determine the exponents numerically\cite{CalculatingLyapunov} (See Appendix \ref{app:Classical Lyapunov}). 
Conservation of phase space volume under Hamiltonian dynamics implies that the exponents (both instantaneous and averaged) sum to zero. Moreover, time-reversal invariance demands that they come in positive and negative pairs. As we shall see below, projection from unitary quantum dynamics to classical dynamics on a variational manifold leads to additional constraints.

\subsection{Quantum Thermalisation} 
\label{QuantumThermo}
There are various ways to express the eigenstate thermalisation hypothesis.
Perhaps the simplest is to state that the expectation of observables should typically have a smooth dependence upon the energy of the state. If this is true for arbitrary states, then it ought to be true for an eigenstate. The expectations of {\it local} operators are the same as in a Gibbs state with the same energy density. 

This locality is crucial. In the conventional view of quantum thermalisation, 
 it allows the major part of the system, over which the observable has no support, to act as a bath for the parts of the system engaged directly in the observation. The remarkable conclusion of the eigenstate thermalisation hypothesis\cite{Deutsch:1991rx,Srednicki:1994qp,Rigol:2008zf} is that the information about thermal averages of local operators is contained in the quantum eigenstates themselves. Coherences in an initial superposition of eigenstates can obscure this fact. Time evolution reveals the underlying thermal properties by a process of dephasing. 
For thermalisation to occur, the part of the system within the observation window must be highly entangled with the system beyond. This is reflected in the fact that states towards the top and bottom of the spectrum --- that in one-dimension are provably weakly entangled\cite{brandao2013area} and suspected to be so in higher dimensions ---  obey the eigenstate thermalisation hypothesis less well than those in the centre\cite{Rigol:2008zf}. 
In this picture, quantum thermalisation depends upon rates of de-phasing, which in turn depend upon differences in the frequencies of the quantum eigenstates of the full system. This is apparently very different from the dependence of classical thermalisation upon the Lyapunov spectrum, although there clearly some relation, since systems that display quantum chaos more rapidly dephase spatial partitions.  Our aim in the following is to further explore the links between classical and quantum thermalisation.

\section{Projecting Quantum to Classical Dynamics}
\label{Sec:ProjectingQuantumDynamics}
Noting the importance of the locality of observation permits an alternative way to frame eigenstate thermalisation that makes much closer contact with its classical counterpart. Central to this is recognising that observations on a spatial partition of a system can be captured accurately by projecting states  to some manifold of variational approximants. Following the projected dynamics on this manifold using the time-dependent variational principle results in a classical Hamiltonian dynamics whose thermalisation is driven by its chaotic properties and characterised by the Lyapunov spectrum. We present two ways to achieve this mapping: approximating the wavefunction of the system using matrix product states (MPS), and approximating the thermofield double purification of the density matrix  by matrix product states. The numerical implementation of these two protocols is very similar --- indeed, we use the same code (with suitable modification) for both cases --- but both their regime of applicability and the manner in which they encode the physics is rather different. 
%
%

\subsection{TDVP applied to the wavefunction}
\label{Sec:TDVPwavefunction}
 A variational parametrisation of a system's wavefunction picks out a sub-manifold of Hilbert space. There are a number of ways that one might choose to project dynamics onto this manifold. The time-dependent variational principle does so by mapping an updated quantum state --- which in general will lie outside of the manifold --- onto the state on the manifold with which it has the highest fidelity. A remarkable property of this mapping is that the reduced equations are those of a classical Hamiltonian system\cite{HaegemanTDVP}. In particular, a quantity conserved by the exact dynamics will also be conserved by the projected dynamics, provided that the symmetry transformation generating it can be captured on the manifold. This permits sensible results to be obtained even at very long times\cite{leviatan2017quantum}.
 
Consider a variational parametrization with a set of complex parameters $\{ X_i \}$. The time derivative of the wavefunction may be written 
$\partial_t |\psi \rangle \approx | \partial_{X_i} \psi \rangle \dot X_i$. It is tempting to substitute this into the Schr\"odinger equation, but the result is  not correct since the action of the Hamiltonian on the state $|\psi({\bf X}) \rangle$ will generally take the state out of the variational manifold. Contracting with a tangent vector $\langle \partial_{\bar X_i} \psi | $ fixes this and permits us to write
\begin{equation}
\langle \partial_{\bar X_i} \psi | \partial_{X_j} \psi \rangle \dot X_j
=
i \langle \partial_{\bar X_i} \psi | \hat {\cal H} | \psi \rangle.
\label{eq:TDVPstate}
\end{equation}
Using a particular basis for the tangent space, one may fix the Gramm matrix $\langle \partial_{\bar X_i} \psi | \partial_{X_j} \psi \rangle = \delta_{ij}$ after which identifying positions and momenta $ q_i  \equiv \sqrt{2}   {\cal I}m X_i $ and $ p_i  \equiv  \sqrt{2}  {\cal R}e X_i $ reduces Eq.(\ref{eq:TDVPstate}) to Hamilton's equation for a classical system\footnote{One may alternatively choose positions and momenta 
$ q_i  \equiv \sqrt{2}   {\cal I}m [M_{ij} X_j] $
 and $ p_i  \equiv  \sqrt{2}  {\cal R}e [M_{ij} X_j] $ 
 with $M^{-2}_{ij}= \bigl< \partial_{\bar X_i} \psi   | \partial_{X_j} \psi \bigr>$
 in order to demonstrate the mapping to a classical system without this explicit gauge fixing.}.  
 Even though the parameters $\{ X_i \}$ may quantify aspects of the entanglement structure of the wavefunction, they nevertheless provide a (semi-) classical description\footnote{This extends the notion of classical chaos considered in Ref.\cite{srednicki1999approach} to semi-classical properties present even without the strict limit of $\hbar \rightarrow 0$.}. 
The technical details of applying this to matrix product states was developed in a seminal work of Jutho Haegeman {\it et al}. We summarise pertinent details in Appendix A. Once the quantum dynamics has been mapped to classical dynamics in this way, we may proceed to evaluation the Lyapunov spectrum associated with this dynamics. 

\subsection{TDVP applied to the thermofield double}
\label{Sec:TDVPthermofield}
As we discuss presently, the MPS ansatz applied in the usual way efficiently describes states near to the top and bottom of the spectrum. States near to the centre of the spectrum we require an alternative variational parametrization. We use an MPS parametrization of the thermofield double. 
The thermofield double\cite{takahasi1974thermo} is a purification of the density matrix. In the eigenbasis of the density matrix $\hat \rho = \sum_\alpha \gamma_\alpha | \alpha \rangle \langle \alpha |$, it may be written as $|\psi \! \! \! \psi \rangle = \sum_\alpha \sqrt{\gamma_\alpha} | \alpha \rangle \otimes | \alpha \rangle$, where $\gamma_\alpha$ are real positive  weights that correspond to the Gibbs weights in thermal equilibrium, and $\alpha$ labels the eigenstates, $|\alpha \rangle$. Physical operators act on the first copy of the state only, so that expectations with the thermofield double are identical to those obtained from the density matrix: $\langle \psi \! \! \! \psi | \hat \theta | \psi \! \! \! \psi \rangle = Tr (\hat \rho \hat \theta)$. The time-evolution of the thermofield double is determined by the Hamiltonian ${\cal H} \! \! \! \!  {\cal H} = {\cal H} \otimes {\bm 1}+{\bm 1} \otimes {\cal H} $, which acts symmetrically on the doubled space. 

Having identified the thermofield double and the appropriate Hamiltonian, we are free to construct an MPS ansatz for it and to evolve using the time-dependent variational principle.  The time-dependent variational principle  projects to the variational manifold by optimising the fidelity of the thermofield double. This amounts to optimising over a certain set of observations --- specifically the trace-norm of the square root of the updated density matrix with the square root of its variational approximation \footnote{The square root guarantees that the fidelity is $1$ for identical density matrices}. The bond order of the MPS for the thermofield double does not have a direct interpretation in terms of the entanglement structure of individual states. Moreover, although evolution under ${\cal H} \! \! \! \! {\cal H} = {\cal H} \otimes {\bm 1}+{\bm 1} \otimes {\cal H} $ without approximation would preserve the purity of a sate,  projection to the variational manifold takes pure states into mixed states. This is consistent with optimising over a certain set of observations, but quite different from the wavefunction MPS which remains pure. Although TDVP has been applied to the density matrix before \cite{2015JChPh.142m4107J}, as far as we are aware, this is the first time that it has been used to follow real time evolution of a matrix product ansatz for it. In order to obtain accurate results, we have made an important modification to the algorithm developed in \cite{HaegemanTDVP} for MPS representations of the state. The MPS for the thermofield double can be written such that the symmetry between the two copies of the physical space is explicit. This is achieved for a bond order $D \!\!\!\! D=D^2$ thermofield MPS by imposing the symmetry $A \! \! \! A^{\sigma \delta}_{I,J}=A \! \! \! A^{\delta \sigma}_{\tilde I, \tilde J}$, where $I\equiv i\otimes i'$ and $ \tilde I \equiv i' \otimes i$ with the indices $i,i',j,j' \in \{ 1,2,... D\}$\footnote{Note that a pure state with bond order $D$ wavefunction MPS  tensor $A^\sigma_{ij}$ can be represented as a thermofield MPS of bond order $D=D\!\!\!\!D$ and tensor $A \! \! \! A^{\delta \sigma}_{i i', j j'}=A^\sigma_{ij}A^\delta_{i'j'}$ }. As described in Appendix A, we employ an additional gauge fixing on the tangent manifold that imposes a constraint gauge equivalent to this. This reduces the number of 
variables and increases the accuracy considerably. 

{\it The Infinite temperature state} takes a particularly simple and instructive form when represented in terms of a thermofield MPS. At $D \!\!\!\! D=1$ it is given by 
$A^{\sigma \delta}= \delta^{\sigma \delta}/\sqrt{2}$. At $D\!\!\!\!D=D^2 >1$ there are many ways to represent the state. A class of symmetrical thermofield MPS states can be constructed from a unitary matrix $U\in SU(dD)$ as
\begin{equation}
A^{\sigma \delta}_{IJ} = \frac{1}{\sqrt{d}} \sum_{\gamma=1}^d U_{(\sigma i),(\gamma j)}U_{(\delta i'),(\gamma j')}.
\label{eq:TInfinity}
\end{equation}
%
This follows from noting that i. the infinite temperature state is the same for any Hamiltonian and ii. that it is invariant under evolution with the Hamiltonian. Eq.(\ref{eq:TInfinity}) follows by representing an arbitrary time evolution of $A^{\sigma \delta}= \delta^{\sigma \delta}/\sqrt{2}$ with a bond operator representation of the time-evolution operator using the unitary $U$. 
%
This manifold of equivalent representations of the infinite temperature state resolves an apparent contradiction: on the one hand a state at the middle of the spectrum of a given Hamiltonian is expected to evolve towards the infinite temperature state, whilst on the other hand the projected dynamics is classically Hamiltonian and so cannot evolve to a single point in phase space. It also holds the seed of how to compress the thermofield MPS representation of a thermalising system at late times\footnote{A. Hallam and A. G Green work in progress.}.

\subsection{Comparing Classical Projections}
These two schemes for projecting quantum dynamics to classical Hamiltonian dynamics capture the physics in rather different ways and have different regimes of validity. The MPS approximation for a state is efficient near the top and the bottom of the spectrum. The bond order required to accurately describe a thermal state at temperature $T$ scales as a double exponential\cite{berta2017thermal}. The thermofield MPS is efficient both at the edges and near to the centre of the spectrum. The latter is justified heuristically as follows: a thermofield MPS of bond order $D\!\!\!\!D$ accurately describes observations up to a lengthscale $\sim \log_{d^2} D\!\!\!\!D$. If this is longer than the thermal correlation length in the final state, the description will accurately capture the dynamics. This occurs near the centre of the spectrum, where the effective temperature is high and the correlation length is short.

These differences are also
revealed in correlation lengths and the factorisation of averages such as $\langle \sigma^x_n \sigma^x_{n+N} \rangle$ for $N$ greater than the thermal correlation length. The wavefunction MPS at low bond order captures such properties in explicit time-averages. The instantaneous correlation length of the wavefunction MPS extracted from its transfer matrix\cite{Orus:2014zl} can be longer than the thermal correlation length, reflecting the long-distance entanglement of its constituent eigenstates. The  thermofield MPS captures the thermal correlation length in a rather different way. Since it is a purification of the density matrix, the thermofield MPS is directly related to observations and already includes the effects of dephasing. In this case, the instantaneous correlation length deduced from the transfer matrix is equal to the thermal correlation length and long distance correlators factorise in instantaneous observations.

\section{Lyapunov Spectrum of Projected Dynamics}
\label{Sec:QuantumLyapunov}
In this section, we summarize how to extract Lyapunov spectra from projected quantum dynamics. The details of this are similar for our two projection schemes. For clarity, we will focus our discussion upon the wavefunction MPS, noting modifications necessary for the thermofield MPS as appropriate. 
\begin{figure}
\includegraphics[width=0.5\textwidth]{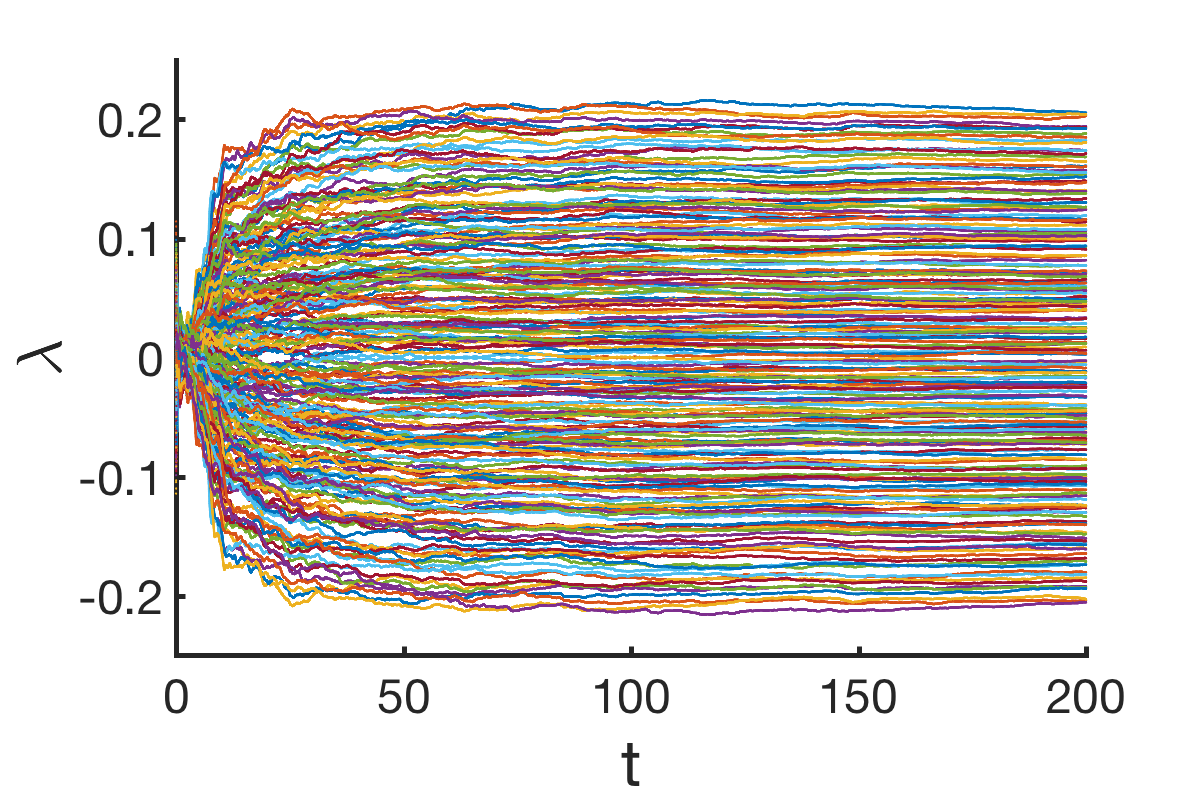}
\caption{
{\it Convergence Plot for a Typical Thermalising System:} The time-averaged Lyapunov exponents {\it versus} time are shown for an MPS representation of the wavefunction of a typical thermalising system. We consider an Ising model with anti-ferromagnetic coupling $J=1$, transverse field $h_x=0.5$ and longitudinal field $h_z=1$. The dynamics are obtained by integrating Eq.(\ref{eq:TDVP}) and the spectrum from averaging instantaneous exponents obtained from Eq.(\ref{eq:lTDVP}) both using bond order $D=10$.}
\label{fig:TypicalConvergence}
\end{figure}

\subsection{Distance on the Variational Manifold}
As a first step to deducing the Lyapunov spectrum, we must assign a distance measure on the variational manifold. This is done using the fidelity between states with two different coordinates ${\bf X}$ and ${\bf X}+d{\bf X}$.  As a simple example, consider a spin-1/2 coherent state given by
$$
|\theta,\phi\rangle = 
e^{-i\phi/2} \cos \frac{\theta}{2} | \uparrow \rangle +e^{i\phi/2} \sin \frac{\theta}{2} | \downarrow \rangle.
$$
The square of the distance between two such states $|\theta, \phi \rangle$ and 
$|\theta+d \theta, \phi+ d\phi \rangle$  can be written, after expanding the fidelity between them to quadratic order, by
$$
dS^2
=1-|\langle \theta,\phi | \theta + d \theta, \phi + d \phi \rangle |^2
=\frac{1}{4}\left( \sin^2 \theta d^2+ \phi d^2\theta \right)
$$
corresponding to the usual distance measure on the Bloch sphere.
In the case of translationally invariant states, we must use the fidelity density rather than fidelity, since the fidelity between translationally invariant states described by an MPS tensor $A^\sigma_{ij}$ and one described by a tensor $A^\sigma_{ij} + dA^\sigma_{ij}$ scales as one over the total length of the system. As described in Appendix A, a small deviation from a translationally invarient MPS state described by a tensor $A^\sigma_{ij}$ may be parametrised\cite{HaegemanTDVP} in terms of freely-chosen complex tensor $X^\sigma_{ij}$: $A^\sigma_{ij} \rightarrow A^\sigma_{ij}+ dA^\sigma_{ij}$.
Surpressing auxiliary indicies for a moment for clarity, we can write 
$dA^\sigma= \sum_{\delta=1}^{d-1} l^{-1/2}V^{\sigma \delta} X^\delta r^{-1/2} $, where $l$ and $r$ are the left and right environments respectively,  and $V^{\sigma\delta}_{ij} \equiv V^{\sigma\delta}_{ij} (A)$ is a tensor of null vectors to ${A^\sigma_{ji}}^*$ (reshaped into a matrix by pairing indices $\sigma$ and $j$). This parametrisation was a crucial development of Haegeman {\it et al} in making the TDVP applied to MPS states tractable\cite{HaegemanTDVP}. 
 The distance measure takes a particularly simple form in terms of $X$:
\begin{equation}
dS^2= \sum_{\sigma i j} {X^\sigma_{ij}}^* X^\sigma_{ji}.
\label{eq:Measure}
\end{equation}
This parametrisation is useful in determining the Lyapunov spectrum, the details of which we turn to next.

\subsection{Linearised TDVP and the Lyapunov Spectrum}
\label{Sec:TDVP}
To extract the Lyapunov spectrum we must characterise the divergence between nearby trajectories. 
Consider two trajectories both in the vicinity of a point on the MPS manifold with tensor $A^\sigma_{ij}$. Let these trajectories have parametrisations in terms of $X^\sigma_{ij}(t)$ and $X^\sigma_{ij}(t)+dX^\sigma_{ij}(t)$, respectively. Substituting each of these into the time-dependent variational principle Eq.(\ref{eq:TDVPstate}) and subtracting, we obtain the following equation for the evolution of the difference between trajectories
\begin{eqnarray}
d \dot X^\sigma_{ij}(t)
&=&
 i \langle \partial_{X^\sigma_{ij}}  \partial_{X^\gamma_{kl}}  \psi | \hat {\cal H} | \psi \rangle dX^\gamma_{kl}(t)
\nonumber\\
& &+
 i \langle \partial_{X^\sigma_{ij}} \psi | \hat {\cal H} | \partial_{X^\gamma_{kl}}\psi \rangle d \bar X^\gamma_{kl}(t).
\label{eq:lTDVP}
\end{eqnarray}
With the minor modification of allowing complex parameters, this equation is clearly analogous to Eq.(\ref{eq:Classical lEoM}) used to extract the Lyapunov spectrum for classical trajectories. Similar structures have been used by Haegeman {\it et al} in order to construct the excitation ansatz\cite{2013PhRvB..88g5133H}, and form the zero-wavevector part of the kernel of a quadratic expansion of MPS path integral about its saddle-point\cite{green2016feynman}. Appendix B gives details of how this equation is evaluated. Extraction of the Lyapunov spectrum now proceeds as in the classical case, using Eq.(\ref{eq:lTDVP}) to find the instantaneous Lyapunov spectrum at each point along a trajectory given by Eq.(\ref{eq:TDVP}) and averaging. 

\begin{figure*}
\includegraphics[width=\textwidth]{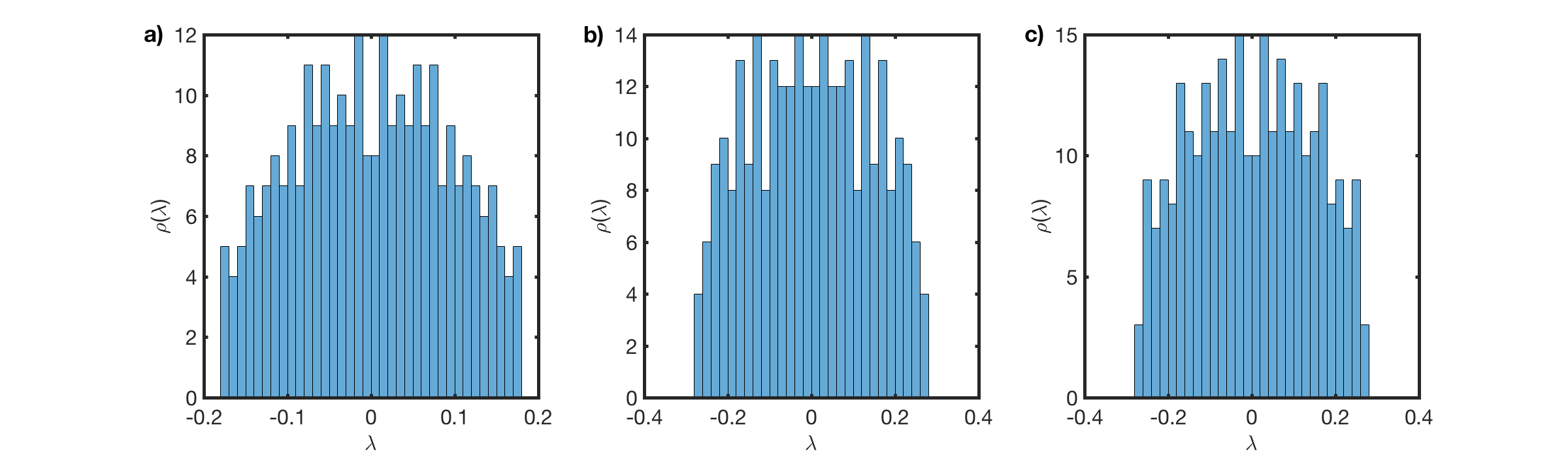}
\caption{
{\it Lyapunov Spectrum for a wavefunction MPS representation of Ising model dynamics}: 
a) Non-integrable case with $J=1$, $h^x=0.5$, $h^z=1$. 
b) Integrable case with $J=1$, $h^x=0.5$, $h^z=0$. 
c) Nearly Integrable case with $J=1$, $h^x=0.5$, $h^z=0.1$ In all cases the spectrum is obtained for an MPS representation of the wavefunction at bond order $D=12$.}
\label{fig:WavefunctionMPSspectra}
\end{figure*}

A final addition to this procedure --- not usually used in calculating Lyapunov exponents for classical dynamical systems ---  is to parallel transport displacements between nearby trajectories along the variational manifold (see Appendix C). This enables us to satisfy some constraints of projected quantum dynamics to numerical precision. The Lyapunov spectra of classical Hamiltonian systems are constrained by time-reversal invariance to have all of the exponents in positive/negative pairs with the same modulus. This property is inherited by the spectrum of projected quantum dynamics. An additional important  property follows from using fidelity to determine the measure on the variational manifold. Fidelity is not changed by unitary time evolution. As a result, Lyapunov exponents calculated for unitary evolution must be identically zero. Evolution under a purely local Hamiltonian provides a useful test case, since it does not change the entanglement structure of a quantum state and the time-dependent variational principle Eq.(\ref{eq:TDVP}) reproduces the full Schr\"odinger equation under projection onto any manifold. The Lyapunov exponents in this case must be identically zero. 
Fig. \ref{fig:TypicalConvergence}  shows typical convergence plots for an MPS approximation to the wavefunction of a thermalising system. The corresponding Lyapunov spectrum is show in Fig. \ref{fig:WavefunctionMPSspectra}.

\section{Numerical Results} 
\label{Results}

In this section, we summarise the results of applying the above methods to the thermalisation of the Ising model with longitudinal and transverse fields:
\begin{equation}
{\cal H} =
\sum_i \left[ J \sigma^z_i \sigma^z_{i+1} + h^z \sigma^z_i+ h^x \sigma^x_i \right].
\label{Ising}
\end{equation}
The properties of this model are well known; it is integrable when the longitudinal field $h^z$ is zero and non-integrable otherwise. This allows us to investigate: i. integrable systems ($J= O(1)$, $h^x= O(1)$ and $h^z=0$), ii. non-integrable/thermalising systems $J= O(1)$, $h^x= O(1)$ and $h^z=O(1)$), and iii. nearly integrable systems $J= O(1)$, $h^x= O(1)$ and $h^z \ll h^x$). We apply the machinery of the time-dependent variational principle to determine trajectories, and the linearised time-dependent variational principle to determine Lyapunov spectra. Reflecting their different encodings of the relevant physics and different regimes of validity, we separate our discussions of the wavefunction MPS and thermofield MPS.

\subsection{Wavefunction MPS}
%
\begin{figure}[b]
\includegraphics[width=0.45\textwidth]{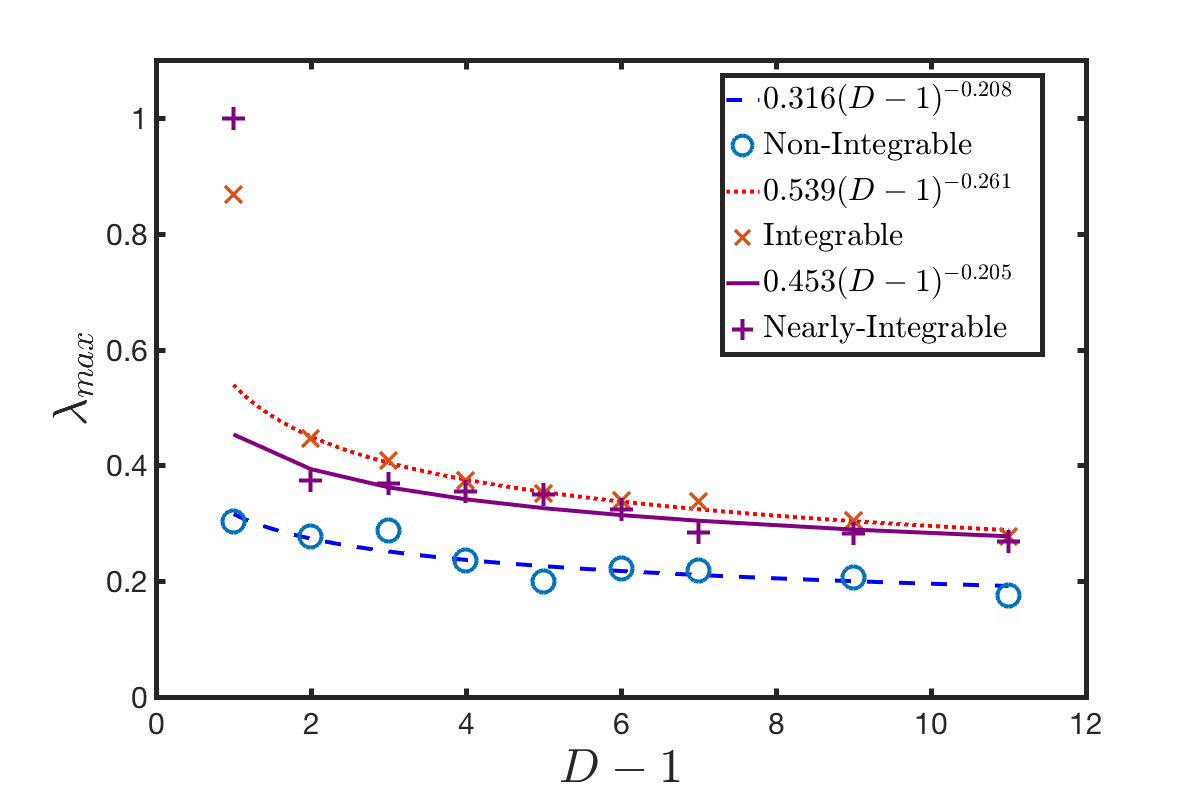}
\caption{
{\it Maximum Lyapunov exponent versus bond order:} The maximum Lyapunov exponent depends strongly upon the projection non-linearities at different bond orders, tending to zero in the limit $D\rightarrow \infty$. Here we show the largest exponent varying with bond order for Non-Integrable (circles), Integrable (crosses) and Nearly integrable (pluses) systems. The largest exponent decreases like $\lambda_{max}(D)=0.316(D-1)^{-0.208}$ for Non-Integrable systems, $\lambda_{max}(D)=0.539(D-1)^{-0.261}$ for Integrable systems and  $\lambda_{max}(D)=0.453(D-1)^{-0.205}$ for Nearly-Integrable systems.  }
\label{fig:ExponentvsBondOrder}
\end{figure}
\begin{figure}[b]
\includegraphics[width=0.45\textwidth]{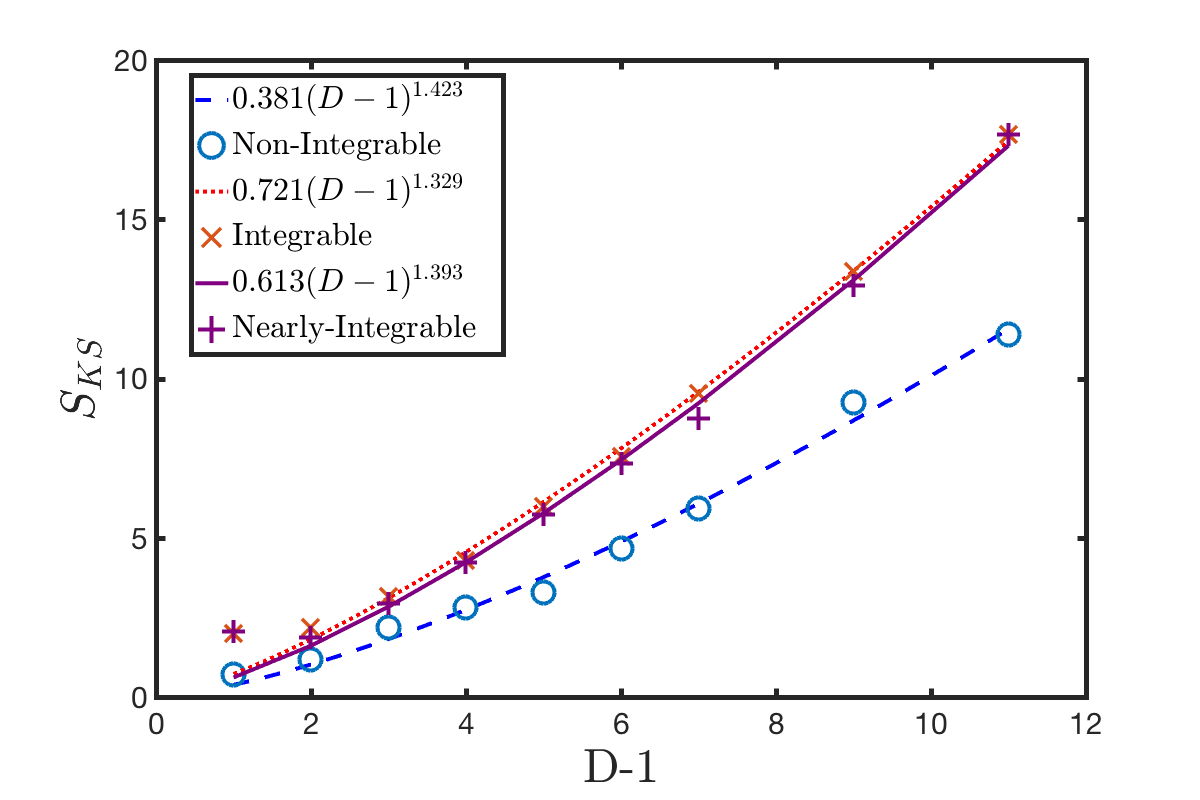}
\caption{
{\it Kolmogorov-Sinai entropy versus bond order:} The Kolmogorov-Sinai entropy is related to entanglement growth at short times, it appears to be diverging with bond dimension.  Here we show the KS entropy varying with bond order for Non-Integrable (circles), Integrable (crosses) and Nearly integrable (pluses) systems. The Non-Integrable KS entropy grows like $S_{KS}(D)=0.381(D-1)^{1.423}$, the Integrable KS entropy grows like $S_{KS}(D)=0.721(D-1)^{1.329}$ and the Nearly-Integrable KS entropy grows like $S_{KS}(D)=0.613(D-1)^{1.393}$. }
\label{fig:KSEntropyvsBondOrder}
\end{figure}
\begin{figure}[b]
\includegraphics[width=0.45\textwidth]{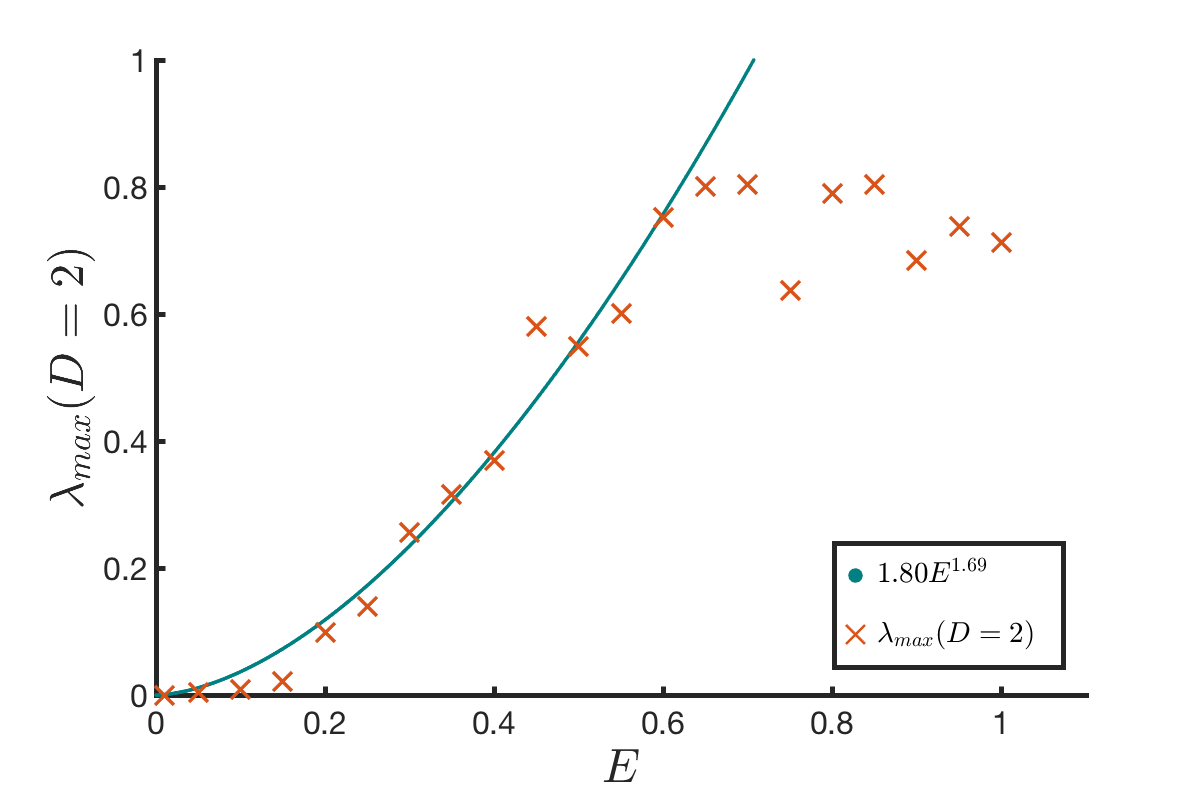}
\caption{
{\it Maximum Lyapunov exponent versus energy density:} It has previously been conjectured that $\lambda_{max} \leq 2\pi k_B T/\hbar$, here  observe that $\lambda_{max}(D=2)$ increases with energy density above the ground state but appears to saturated at $E\approx0.6$. }
\label{fig:ExponentvsEnergy}
\end{figure}

\begin{figure}[t]
\includegraphics[width=0.45\textwidth]{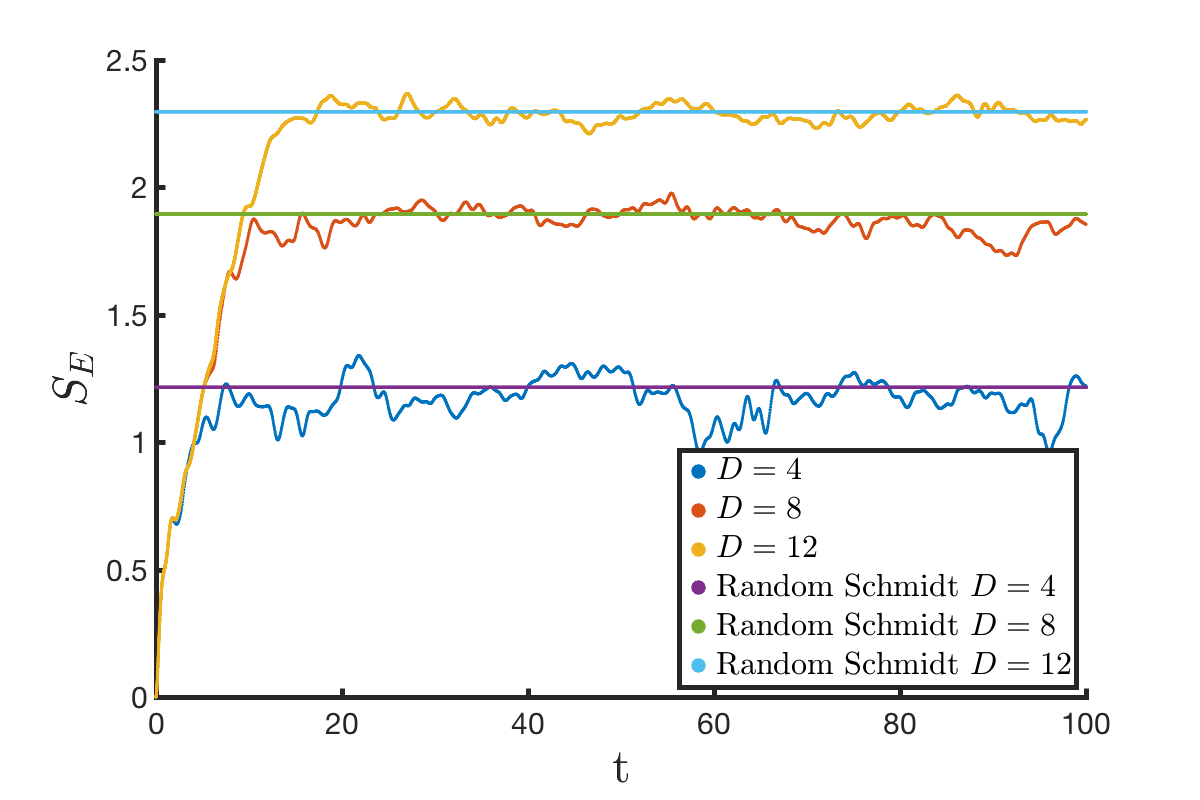}
\caption{
{\it Entanglement entropy across a bond compared to randomly distributed Schmidt coefficients:} At a given bond dimension the entanglement entropy will saturate after a short time. The saturation value for the entanglement entropy is in strong agreement with a random uniform distribution of Schmidt coefficients as discussed in the text. }
\label{fig:Entropy_vs_randomschmidt}
\end{figure}
\begin{figure}
\includegraphics[width=0.45\textwidth]{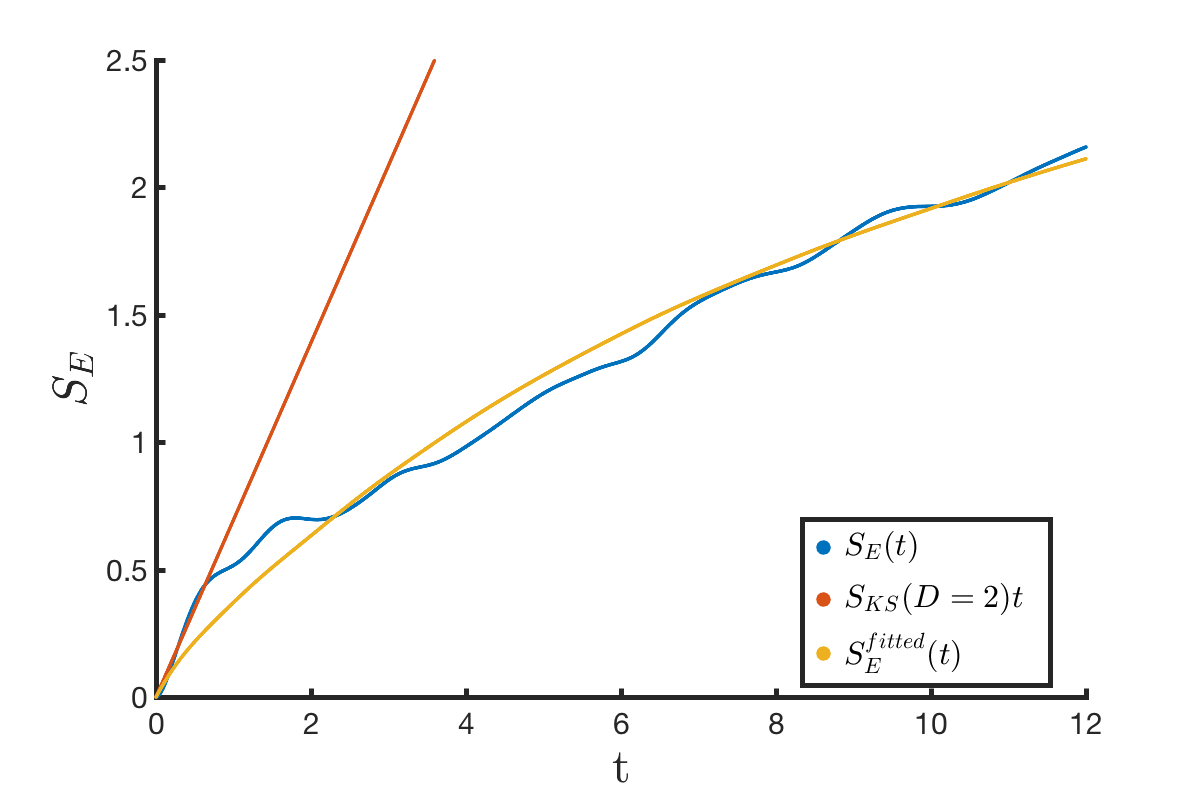}
\caption{
{\it Entanglement entropy and Kolmogorov-Sinai entropy:} The gradient of the entanglement entropy is determined by the Kolmogorov-Sinai entropy. Here we demonstrate the short time behaviour of the entanglement entropy is determined by $S_{KS}(D=2)$.  By integrating  Eq.(\ref{eq:SKGeneral}) after substituting $D(t)$ from Eq.(\ref{eq:Doft}) and the fitted form of $S_{KS}(D)$ from Eq.(\ref{eq:SKfit}) we find a  zero-parameter fit between the Lyapunov spectrum and entanglement.}
\label{fig:Entanglement_Dt}
\end{figure}
\begin{figure*}
\includegraphics[width=\textwidth]{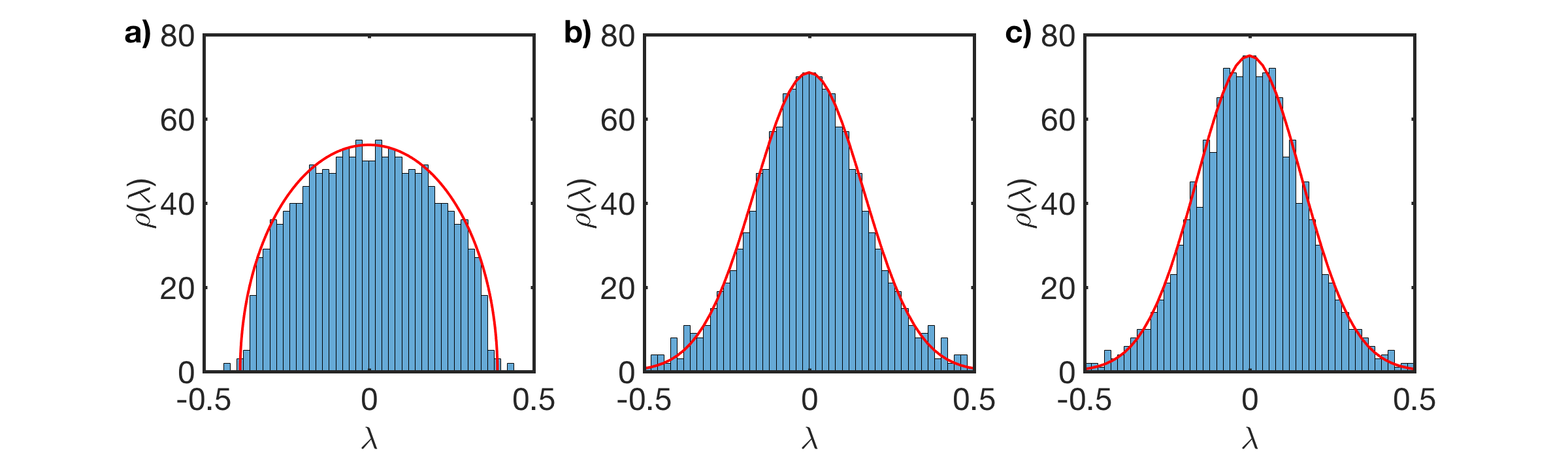}
\caption{
{\it Lyapunov Spectrum for a thermofield MPS respresentation of Ising model dynamics}:
a) Non-integrable case with $J=1$, $h^x=0.5$, $h^z=1$. 
b) Integrable case with $J=1$, $h^x=0.5$, $h^z=0$. 
c) Nearly Integrable case with $J=1$, $h^x=0.5$, $h^z=0.1$ In all cases the spectrum is obtained for  a wavefunction MPS at bond order $D \!\!\!\!D=16$. The non-Integrable case appears to fit a semicircle distribution with radius $r=0.39$, the Integrable case appears to be Gaussian with standard deviation $\sigma=0.167$ and the nearly integrable case appears to be Gaussian with standard deviation $\sigma=0.161$.}
\label{fig:ThermofieldMPSspectra}
\end{figure*}

We now consider the Lyapunov spectra evaluated from the wavefunction MPS starting from an initial product state
$\ket{\psi(0)}_i=  (0.382 - 0.382i)\ket{\uparrow}_i+(-0.595 + 0.595i)\ket{\downarrow}_i$ near the bottom of the spectrum. The Lyapunov spectrum for the non-integrable, integrable and nearly-integrable cases are shown in  Fig. \ref{fig:WavefunctionMPSspectra}.
All show a broad distribution of exponents, with no strong differences apparent between integrable and non-integrable cases. 

\vspace{0.1in}
\noindent
{\it Spectrum vs Bond Order:}
Since the nonlinearities and chaos of our dynamics arise from projection to the variational manifold, the Lyapunov spectrum varies with bond order. This situation is unlike the conventional use of matrix product methods, where increasing bond order give increasingly accurate results.
The dependence of the maximum Lyapunov exponent, $\lambda_{max}$, with $D$ is shown in Fig \ref{fig:ExponentvsBondOrder}. This appears to show a monotonic decrease from $D=2$ as $ D\rightarrow \infty$. Note that in the translationally invariant case with spin $1/2$, the projected dynamics is not chaotic at  $D=1$ by the Poincar\'e-Bendixson theorem, since the phase space is two-dimensional. 
The following discussion demonstrates the consistency of these results with physical observations.

Maldacena et al.\cite{2016JHEP...08..106M} have conjectured that the largest Lyapunov exponent of a quantum system has an upper bound related to its temperature $\lambda_{max} \leq 2\pi k_B T/\hbar$ . The behaviour of $\lambda_{max}$ for initial states of different energy can be seen in Fig \ref{fig:ExponentvsEnergy}. At low energies the exponent appears to increase as a power law before saturating at $E\approx0.6$.

\vspace{0.1in}
\noindent
The dependence of the {\it Entanglement Entropy}, $S_E$ upon time is shown in Figs.~\ref{fig:Entropy_vs_randomschmidt} and \ref{fig:Entanglement_Dt}. For a given bond order, $S_E$ saturates. To a good approximation (that we can manipulate analytically) this saturation value corresponds to drawing the Schmidt coefficients $s_n$ from a distribution given by the modulus of the elements of a random $O(D)$ vector. The mean Schmidt coefficients then correspond to 
$s_n =n  \sqrt{6}/\sqrt{D(1+D)(1+2D)} $,
from which one may deduce a saturation entanglement at large bond order given by
\begin{equation}
S^{\hbox{Sat}}_E(D)= - \sum_{n=1}^D s^2_n \log s^2_n \approx \log \left(\frac{D e^{2/3}}{3} \right).
\label{SaturationEntanglement}
\end{equation}
With growing entanglement, the effective bond order of the quantum state (the bond order required for an accurate description) grows. We can use Eq.(\ref{SaturationEntanglement}) to deduce the time-dependence of the bond order; $D$ attains particular value at point where $S_E(t)$ crosses the corresponding saturation value. A continuous approximation can be found by equating 
$S^{\hbox{Sat}}_E(D)= S_E(t) $, 
from which we obtain 
\begin{equation}
D(t)= 3 e^{-2/3} e^{S_E(t)}
\label{eq:Doft}
\end{equation}
in the large bond-order limit. This dependence of bond order upon time allow us to demonstrate the consistency of the Lyapunov spectrum and its variation with $D$ with the physically relavent dependence of the entanglement entropy upon time. 

\vspace{0.1in}
\noindent
The {\it Kolmogorov Sinai} entropy is a measure of how quickly knowledge of a system's initial state is lost in a chaotic system. It determines the rate of growth of the volume (of gyration) of a region of phase space and, following Pesin's theorem\cite{pesin1977characteristic}, is given by the sum of the positive Lyapunov exponents. Fig.\ref{fig:KSEntropyvsBondOrder} shows the Kolmogorov Sinai entropy calculated from our Lyapunov spectra and its dependence upon bond order. The latter dependence is fitted in the non-integrable case with a polynomial approximation\footnote{Note that the expansion is in $D-1$ since the Lyapunov exponents and so $S_{KS}$ are zero at $D=1$. }
\begin{equation}
S_{KS}(D) =0.381 (D-1)^{1.423}.
\label{eq:SKfit}
\end{equation} 
Studies of single particle quantum chaos have shown the relationship $\dot S_E(t=0) = S_{KS}$, provided that starting wavefunction is as classical as possible\cite{1995PhyD...83..300Z,rozenbaum2017lyapunov,rozenbaum2018universal}. Here we find  --- as indicated in Fig.~\ref{fig:Entanglement_Dt} --- that 
$\dot S_E(t=0)= S_{KS}(D=2)$. $D=2$ corresponds to the most classical, non-trivial (recall that $D=1$ has vanishing Lyapunov exponents) projected dynamics and is the many-body equivalent of the single particle result. We speculate the following extension of this result:
\begin{equation}
\dot S_E(t) = \frac{S_{KS}(D(t))}{(D(t)-1)^2}.
\label{eq:SKGeneral}
\end{equation}
Our main justification for this is the very good, zero-parameter fit that it gives between our results for the entanglement and Lyapunov spectrum. 
Fig.~\ref{fig:Entanglement_Dt} shows the time-integral of the right-hand side of Eq.(\ref{eq:SKGeneral}) substituting $S_{KS}(D)$ from Eq.(\ref{eq:SKfit}) and $D(t)$ from Eq.(\ref{eq:Doft}), alongside the entanglement entropy. 

Eq.(\ref{eq:SKGeneral}) can also be used to place bounds upon the Lyapunov spectrum. At long times we expect $S_E(t) \sim t$ for thermalising systems. Assuming the validity of Eq.(\ref{eq:SKGeneral}) then, $S_{KS}(D) =\alpha D^2$ at large $t$ suggesting that the fit in Eq.(\ref{eq:SKfit}) must be modified at large $D$.  Moreover, if the exponents converge to a consistent distribution then this also implies that $\lambda_{max}$ converges. For example, if the exponents approach a semicircular distribution then $\pi \lambda_{max}^2(D)/4=\alpha$ to be consistent with $S_{KS}(D) = \alpha D^2$.

It is apparent from these observations that the Lyapunov spectrum extracted from mapping the quantum dynamics of the wavefunction to classical Hamiltonian dynamics is not unique. Although the dependence is rather slow with bond order, there is no sense in which spectra collected in this way show numerical convergence, for example with increasing bond order. A moments reflection about the way in which the wavefunction MPS captures the physics of thermalisation shows why. At low bond order, the dynamics is very non-linear and thermalisation occurs {\it via} chaotic classical dynamics. Thermal averages are recovered in temporal averages of the simulated dynamics. 
As bond order increases, the MPS ansatz make better and better approximation to the underlying eigenstates and ultimately, thermalisation is captured in the same way as the conventional picture of eigenstate thermalisation. Thermal averages are obtained in instantaneous measurements after an initial period of dephasing reveals the intrinsic properties of the underlying eigenstates.  However, the Lyapunov spectrum does have physical meaning. We have demonstrated how the physical quantity, $S_E(t)$, is related to the Lyapunov spectrum obtained on a  variational manifolds.

\subsection{Thermofield MPS}

\begin{figure}[b]
\includegraphics[width=0.45\textwidth]{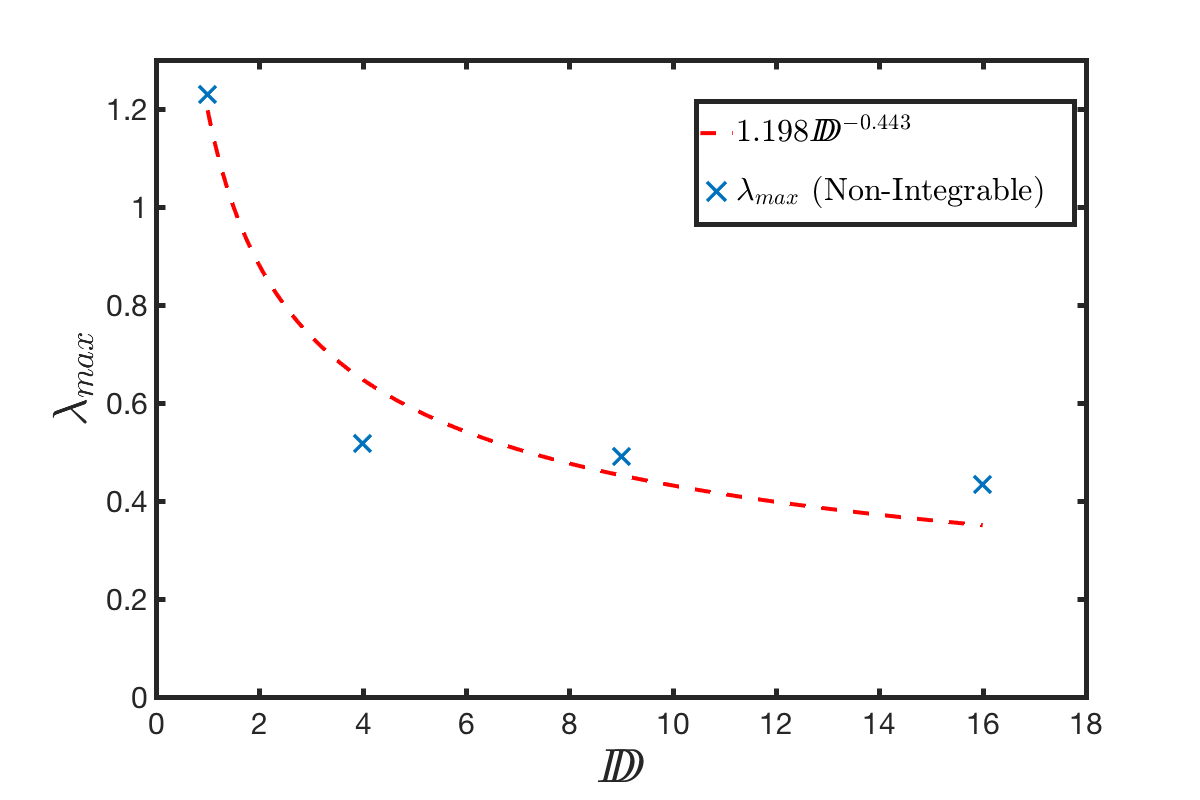}
\caption{
{\it Maximum Lyapunov Exponent vs Thermofield MPS Bond Order for Non-Integrable System:} The largest Lyapunov exponent for the Ising model with $J=1$, $h^x=0.5$, $h^z=1.0$ obtained for an MPS representation of the Thermofield double. The exponent appears to be approaching zero like $\lambda_{max}=1.198 D\!\!\!\!D^{-0.443}$  }
\label{fig:lambdamaxvsD_thermal}
\end{figure}
\begin{figure}[b]
\includegraphics[width=0.45\textwidth]{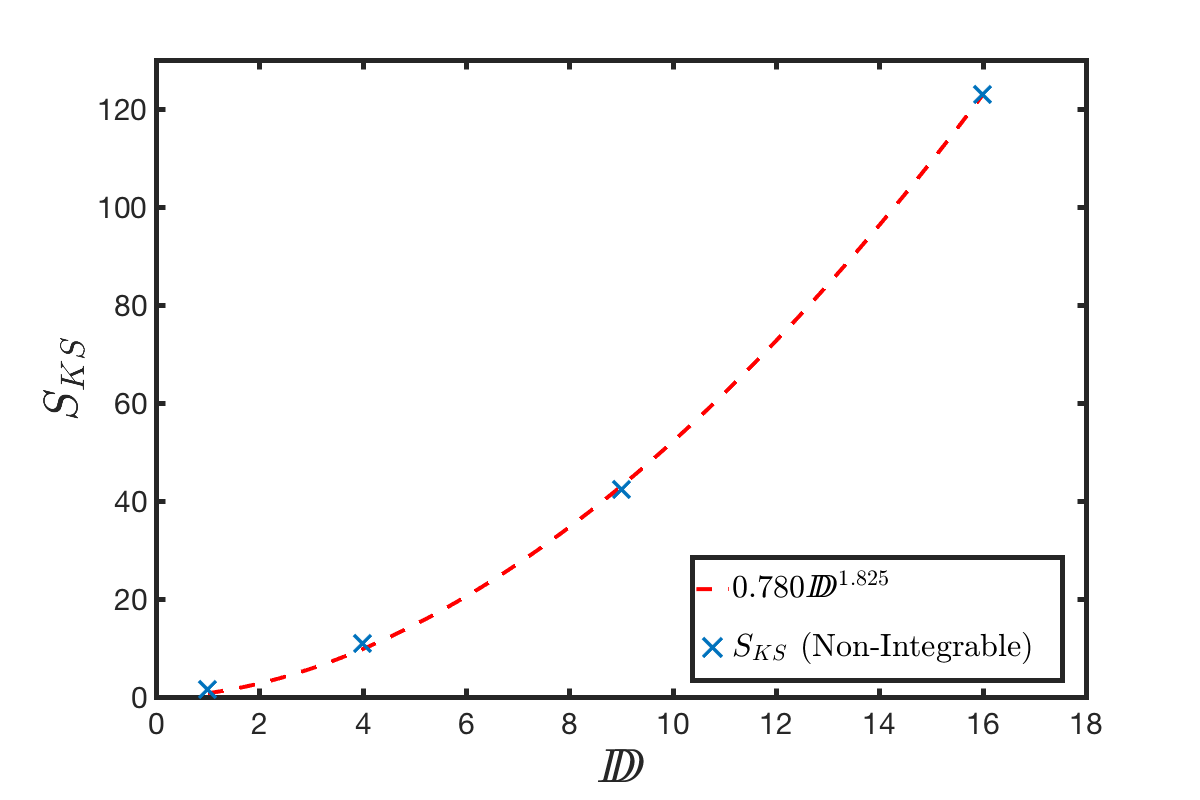}
\caption{
{\it Kolmogorov-Sinai entropy vs Thermofield MPS Bond Order for Non-Integrable System:} The Kolmogorov-Sinai entropy for the Ising model with $J=1$, $h^x=0.5$, $h^z=1.0$ obtained for an MPS representation of the Thermofield double. The KS entropy appears to be divering, growing like $S_{KS}=0.780D\!\!\!\!D^{1.825}$. }
\label{fig:KSentropy}
\end{figure}
\begin{figure}
\includegraphics[width=0.5\textwidth]{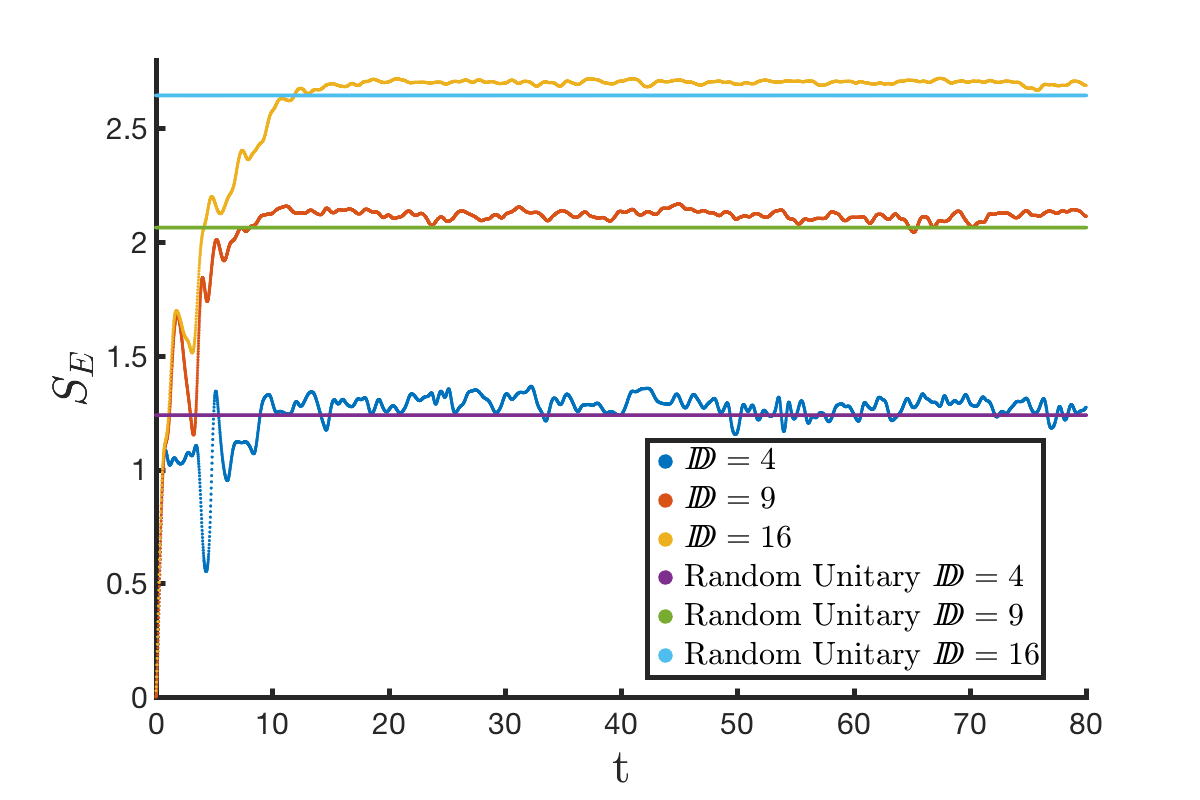}
\caption{
{\it Entanglement of the midspectrum state:} The entanglement between sites of a Thermofield MPS state starting in the middle of the spectrum saturates at a value close to value obtained by averaging over random unitaries. }
\label{fig:Thermalizingstate_vs_randomunitary}
\end{figure}

The above analysis allows us to related the chaos of projected quantum dynamics near to the edge of the spectrum to the process of thermalisation. As discussed in Sec.~\ref{Sec:ProjectingQuantumDynamics} a matrix product state description of the wavefunction cannot work near to the centre of the spectrum. In this subsection, we apply our analysis of the Lyapunov spectrum to a matrix product state representation of the thermofield double. We consider an initial pure state near to the middle of spectrum, $\ket{\psi(0)}_i=  0.448\ket{\uparrow}_i+0.873\ket{\downarrow}_i$. The late time dynamics of this are similar to the infinite-temperature state. 

{\it The Lyapunov spectra for the thermofield MPS} dynamics are shown in Fig.~\ref{fig:ThermofieldMPSspectra}. There is a clear distinction between the non-integrable, and integrable and nearly-integrable cases. The former has a semi-circular distribution, whereas the latter are narrower and  fit a Gaussian distribution (with long tails that have been cut off in Fig.~\ref{fig:ThermofieldMPSspectra}).  The semi-circular distribution in the non-integrable case suggests a connection to random matrix theory. Such a connection has previously been explored in the context of quantum gravity \cite{2016JHEP...02..091G,2018PhRvE..97b2224H}. 

Fig. \ref{fig:lambdamaxvsD_thermal} shows the variation of the {\it maximum Lyapunov exponent} with bond order. The symmetry constraint that we impose upon the thermofield MPS tensor restrict the bond order to $D \!\!\!\!D=1,4,9,16$  {\it etc} \footnote{NB: since the  dimension of the local Hilbert space is $d^2$, dynamical chaos occurs at $D\!\!\!\!D=1$.}, and together with the rapid growth of the number of Lyapunov exponents as $2(d^2-1) D\! \! \!\!D^2$ this leads to rather few points in the figure. Our numerics are fit by $0.572 D\!\!\!\!D^{-0.045}$, or $0.567e^{-0.0173 D\!\!\!\!D}$, but are also consistent with convergence  $0.410+0.1740 e^{-0.0116 D\!\!\!\!D}$. The latter might be expected since the thermofield double (being a purification of the density matrix) encodes a limited set of observations corresponding roughly to a window of size $\frac{1}{2} \log_2 D \!\!\!\! D $. When this window is larger than the correlation length timescales of the dynamics are expected to converge to values characteristic of the observable thermalisation process\footnote{In the limit $D \!\!\!\! D \rightarrow \infty$ exponents should still tend to zero. Since the spectrum is discrete, however, there is room for some to remain finite.}

{\it The Kolmogorov Sinai } entropy for the thermofield MPS is shown in Fig. \ref{fig:KSentropy}. This is fit with $0.780 D\!\!\!\!D^{1.825}$ to high accuracy. This scaling is less than $D\!\!\!\!D^2$ (the voulme of phases space) of a typical classical dynamical system. This is consistent with unitary dynamics as $D\!\!\!\!D$ tends to infinity. {\it The thermofield entanglement} is shown in Fig.~\ref{fig:Thermalizingstate_vs_randomunitary}. At short times the thermofield time evolution is exact and the entanglement is double the matrix-product entanglement. When the thermofield state becomes mixed the thermofield entanglement appears to be more closely related to operator entanglement\cite{zanardi2001entanglement,zhou2017operator}. It is interesting to note that the saturation of this thermofield entanglement is near to the mean value obtained by averaging the thermofield entanglement of the infinite-temperature state given by Eq.(\ref{eq:TInfinity}) with a Haar measure over $U$. Finally, we note that unlike wavefunction MPS, we have not been able to determine an simple relationship between the Kolmogorov-Sinai entropy and the thermofield entanglement.

\section{Discussion}
The analysis presented above allows the thermalisation of local observables to be recast as a chaotic classical Hamiltonian dynamics in two different ways: using the time-dependent variational principle to evolve MPS representations of the wavefunction and of the thermofield double. This picture is complementary to the dephasing of eigenstates in the conventional picture of eigenstate thermalisation and brings the study of quantum chaos full circle. Early studies of quantum chaos focussed upon single particle quantum systems whose semi-classical limit is chaotic. The impact of this upon the level statistics provides a convenient way to discriminate between chaotic and non-chaotic behaviour that can be extended to many-body systems. Our approach returns to a semi-classical analysis for many-body systems.  Albeit, the semi-classical dynamics that we study describes entanglement structure whose origin is quantum mechanical. We have applied this to the Ising model with a longitudinal and transverse field using  the time-dependent variational principle applied to matrix product states.

This analysis has afforded several insights. An MPS description of the wavefunction suggests a new relationship between the Komogorov-Sinai entropy (and its dependence upon bond order) and the entanglement, Eq.(\ref{eq:SKGeneral}). This relationship holds not just for the initial entanglement growth, but rather for the entire time-dependence of the entanglement. Using the thermofield MPS reveals a semi-circular distribution of Lyapunov exponents in the non-integrable case and Gaussian distribution in the integrable case. The former result has been anticipated in the context of gravitation\cite{2016JHEP...02..091G,2018PhRvE..97b2224H}, where it was hypothesised that it may be universal. 

 There are several natural extensions of the present work. 
 Similar Lyapunov spectra calculated for finite systems would enable comparison
calculations of the out-of-time ordered correlator. The latter has become a canonical tool for studying quantum thermalisation\cite{2017NJPh...19f3001B,2017arXiv171009827R,2018PhRvX...8b1013V,2017AnP...52900332C}. When studying finite systems it may be more convenient to calculate Lyapunov exponents using a time-series approach \cite{WOLF1985285,1991PhRvA..43.2787B,1994PhLA..185...77K,PhysRevA.34.4971}. This would involve extracting exponents from the evolution of observables, it is currently unclear if exponents can be accurately calculated in the quantum context using this approach.

{\it Effective Long-time Dynamics:}
The exponential increase of data required to accurately describe the dynamics of a quantum system at early times and the ultimate decrease of this data at late times for thermalising quantum systems presents an acute difficulty for efficient numerical simulation. Whilst the mechanism of this decrease can be understood by dephasing, it is difficult to turn this insight into a practical scheme. The new perspective provided here might provide a route. The dynamical modes  of a classically chaotic system divide naturally into those that have revealed their chaotic nature on a given timescale and those that have not. The latter behave as quasi-regular modes and the former as a chaotic bath for them. This division, suggests an appealing way to describe late-time dynamics of the wavefunction MPS. On the longest timescales, the majority of modes form a bath, with energy density equal to that of the initial state. It ought to be possible to develop a Langevin description of this late-time dynamics. The cross over between early- and late-time dynamics being captured as the crossover from dominance of inertial dynamics to diffusive dynamics driven by the noise and dissipation. Such a description may be developed\footnote{Using the recently developed path integral over matrix product states\cite{green2016feynman,kamenev2011field} from a Keldysh path integral, a Langevin equation can be constructed in the usual way.} by adding noise and dissipation to the time-dependent variational principle, to derive an MPS Langevin equation. This picture provides a suggestive link to random circuit analyses of thermalisation\cite{brown2012scrambling,cramer2012thermalization,vznidarivc2012subsystem,hamma2012quantum,brandao2012convergence,nahum2017quantum,von2018operator,jonay2018coarse,nahum2018operator}. 

However, our implementation of the time-dependent variational principle for the evolution of a thermofield MPS may obviate the need for such a Langevin description. Eigenstate thermalisation suggest that there should be an efficient description of both early and late time dynamics. If a single variational scheme can capture both limits --- and if it is imbued with sufficient variables to surmount the information barrier in the middle of the dynamics --- then it should be possible to obtain an accurate numerical description that runs from the earliest times to the latest times. The time-dependent variational principle applied to the thermofield MPS seems to satisfy these requirements. The remaining ingredient is to find a way of compressing the thermofield description at late times. As commented in Sec.~\ref{Sec:TDVPthermofield}, the multiple, equivalent descriptions of an infinite temperature state contains the essence of  such a compression. Implementing this is a subject of ongoing investigation.

{\it Glimmers of a Quantum KAM Theorem:\cite{Konik:2015hm}}
Classical integrable systems show a remarkable robustness to perturbation. The KAM theorem shows that aspects of integrability remain through the presence of residual invariant tori (essentially periodic motions of action angle variables) when perturbations away from integrability are below some threshold. There has been speculation recently of whether such effects could be apparent in a quantum system\cite{Konik:2015hm}. It is {\it inevitable} that they are possible when quantum dynamics is projected to (semi-)classical dynamics by observing on a finite window. This is a promising direction for future study,  for example in the study of many body localisation.

{\it Thermalisation in Quantum Critical Systems:}
Matters of thermalisation and chaotic dynamics come to a head in quantum critical systems. These are the most rapidly dissipating and de-phasing of quantum systems\cite{zaanen2004superconductivity}, and it is no coincidence that recent years has seen their mapping to black holes --- through the AdS/CFT correspondence --- themselves the most rapidly scrambling (classically chaotic) of objects\cite{Susskind:2008sp}. The semicircle distribution of Lyapunov exponents that we have uncovered already makes links to works carried out in this context\cite{2016JHEP...02..091G,2018PhRvE..97b2224H}. A direct application of MPS methods has limitations for the study of quantum criticality, however, because of diverging correlation lengths. It may be that other variational schemes such as MERA can do a better job, although in that case, dynamics are trickier. The view of quantum dynamics that we present should, then give an interesting complementary view of dynamical transitions observed after quenches and sweeps through quantum critical points. 

To conclude, we have presented techniques that provide a bridge between thermalising quantum systems and classically chaotic Hamiltonian systems. Moreover, we have demonstrated how techniques developed in the latter may be applied fruitfully to the study of quantum thermalisation.  We hope that this approach will provide a useful insights into other aspects of quantum thermalisation and chaos. As anticipated by many others, the intrinsic chaos of non-linear classical mechanics is the very property that permits the stability of a classical view of the underlying quantum world. 

\bibliography{bibliography}

\appendix
\section{Extracting Lyapunov Exponents}
\label{app:Classical Lyapunov}
Here we provide some additional details of how to extract Lyapunov spectra from linearised equations of motion, Eq.(\ref{eq:Classical lEoM}), describing the evolution of the displacement between neighbouring trajectories ${\bf X}(t)$ and ${\bf X}(t)+d{\bf X}(t)$.
The asymptotic rate at which these two trajectories diverge (or converge) is characterized with a Lyapunov exponent.  If the solution for this equation is $d {\bf X}(t)=Y({\bf X},t)d {\bf X}(0)$ then the Lyapunov exponent associated with these trajectories is
\begin{equation}
\lambda  =\lim_{t \rightarrow \infty} \frac{1}{t}\log{\frac{|d {\bf X}(t)|}{|d {\bf X}(0)|} } = \lim_{t \rightarrow \infty} \frac{1}{t}\log{(  Y({\bf X},t)d {\bf X}(0) )}.
\label{eq:Max Lyapunov}
\end{equation}
For almost all trajectories ${\bf X}(t)$ and almost all tangent vectors  $d{\bf X}(t)$ the limit in Eq.(\ref{eq:Max Lyapunov}) converges to the largest Lyapunov exponent of the system \cite{oseledec1968multiplicative,ruelle1979ergodic}. 

Using a similar approach it is possible to calculate the entire Lyapunov spectrum. Instead of a single trajectory, consider a $d$-dimensional parallelepiped defined by $d$ vectors tangent to the manifold at point ${\bf X}(t)$, ${\bf U}(t)=\{d {\bf X}^1(t), d{\bf X}^2(t),..., d{\bf X}^d(t)\}$. The volume of the parallelepiped will evolve over time in a manner determined by the  $d$ Lyapunov exponents 
\begin{equation}
\sum^d_{i=1} \lambda_i =\lim_{t \rightarrow \infty} \frac{1}{t}\log{( \text{Vol}^d (Y({\bf X},t) {\bf U}(0)) )}.
\label{eq:Lyapunov parallelepiped}
\end{equation}
Unfortunately, the Lyapunov spectrum cannot be easily extracted using this method. As $t\rightarrow \infty$ the different tangent vectors comprising the parallelepiped all begin to point in the direction of the largest Lyapunov exponent. Many methods have been introduced to circumvent this issue. We use an algorithm introduced by Bennetin et al. \cite{1980Mecc...15....9B}. 

An orthonormal basis for the tangent space ${\bf V}(t)=\{d {\bf \hat{X}}^1(t), d{\bf \hat{X}}^2(t),..., d{\bf \hat{X}}^d(t)\}$ is defined and then evolved for a short time:
\begin{equation}
{\bf U}(t+\delta t)=     Y({\bf X},t){\bf V}(t).
\label{eq:Tangent Space Evo}
\end{equation}
This evolution rotates and changes the length of each of the unit vectors in ${\bf V}(t)$. By performing  a QR decomposition on ${\bf U}(t+\delta t)$ we can separate these two effects:
%
${\bf U}(t+\delta t) = {\bf Q}(t+\delta t) {\bf R}(t+\delta t)$ .
%
${\bf V}(t+\delta t) \equiv {\bf Q}(t+ \delta t)$ is a new orthonormal basis for the tangent space, obtained by rotating the basis vectors from the previous time step. Since $\text{det}[U(t+\delta t)]=\prod_{i}R_{ii}$ the diagonal elements of ${\bf R}(t+\delta t)$ capture the extent to which the volume of the parallelepiped at the previous time step has changed. 

This process is repeated iteratively to obtain a sequence of matrices ${\bf R}(t)$ from which me may extract the Lyapunov spectrum using 
%
%
\begin{equation}
\lambda_i =\lim_{N \rightarrow \infty}\frac{1}{N \delta t} \sum^N_{n=1}\log{|R_{ii}(n\delta t)|}.
\label{eq:Lyapunov}
\end{equation}


\section{Implementing the TDVP}
\label{app:TDVP}
Here we provide details of the time-dependent variational principle used to generate our numerical results. Our implementation closely follows that of Haegeman {\it et al}\cite{HaegemanTDVP} and we refer to the original papers for further details. Here we give a brief summary noting in particular aspects that require modification for the thermofield MPS.

\subsection{Matrix product state TDVP}
\label{App:TDVP}
A variational wavefunction $\ket{\psi( A)}$ defined  by a matrix product state $  A^\sigma_{ij}$ evolves on the manifold of matrix product states according to Eq.(\ref{eq:TDVPstate}) with the appropriate identification of varaibles and indices: $X \rightarrow A$, $i \rightarrow I \equiv \{ i,j,\sigma \}$ giving 
\begin{equation}
\langle \partial_{\bar A_I} \psi | \partial_{A_J} \psi \rangle \dot A_J
=
i \langle \partial_{\bar A_I} \psi | \hat {\cal H} | \psi \rangle.
\label{eq:TDVP}
\end{equation}
Determining the time evolution of $\ket{\psi(A)}$ from equation Eq.(\ref{eq:TDVP}) requires inversion of the Gram matrix $\langle \partial_{\bar A_I} \psi | \partial_{A_J} \psi\rangle$. In the case of matrix product states this is a $d D^2 \times d D^2$ matrix, however not all of the $dD^2$ tangent vectors are linearly independent so the Gram matrix cannot be inverted. As noted in Ref.\cite{HaegemanTDVP}, this
can be resolved by imposing a gauge fixing condition on the states $\ket{\partial_{A_J}  \psi(A)}$ parameterizing the tangent space.  We follow Ref.\cite{HaegemanTDVP} and use the \emph{left tangent gauge fixing condition},
%
$\sum_{\sigma=1}^d A_{ij}^{\sigma \dagger} l_{jk} dA^{\sigma}_{kl}=0$,
%
where $l$ is the left environment, {\it i.e.} the result of contracting the MPS state with its conjugate on every site to the left of a given site and $dA$ is an update to the MPS tensor such that $A \rightarrow A + dA$. 
This gauge condition can be achieved by constructing $L_{i,(\sigma j)}=[A^\sigma l^{\frac{1}{2}}]_{i j}$ and calculating its null vectors, $[V_L])_{(i \sigma), j}$. If the null space is reshaped to $[V_L]^\sigma_{ij}$ then a $dA$ that satisfies the tangent gauge condition can be written as 
\begin{equation}
dA^\sigma(x)= l^{-\frac{1}{2}}V^\sigma_LXr^{-\frac{1}{2}},   
\label{eq:tangent space parameterization}
\end{equation}
where $r$ is the right environment..
Using this parameterizing the Gram matrix becomes diagonal and the time evolution of the state can determined by evaluating Eq.(\ref{eq:TDVPstate}) to find the $(d^2-1)D \times D$ matrix $X$. 

\subsection{Inverse-free algorithm}
While this algorithm is sufficient to determine the time evolution of a matrix product state at fixed bond dimension it has two flaws. Firstly, it necessarily involves inverting Schmidt coefficients and therefore encounters issues when a state has small Schmidt values. 
Secondly, there is no easy way to increase the bond dimension of the matrix product state as we may need to if we start from say a product state initial state. Both of these issues can be solved by using an inverse-free version of the TDVP algorithm\cite{haegeman2016unifying}. Here we provide minor modifications to this algorithm required to study real-time evolution rather than imaginary-time evolution as studied in \cite{haegeman2016unifying}. 

An inverse-free algorithm uses $A$ in both left and right canonical forms, $A_L$ and $A_R$ respectively. For $A_L$ the dominant left eigenvector of the transfer matrix is $l=\mathbb{I}$ and dominant right eigenvector is $r=c c^\dag$. For $A_R$ the dominant left eigenvector of the transfer matrix is $l=c^\dag c$ and dominant right eigenvector is $r=\mathbb{I}$. The algorithm has three key steps:

\noindent
i. $A_R$ and $c$ can be calculated from $A_L$ in an inverse-free method by iterating 
 \begin{equation}
     [c_{(i+1)},A_{R(i+1)}]=RQ(A_Lc_i)
 \label{eq:A_R from A_L}
 \end{equation}
 until $c_{i+1}\approx c_i$ where $RQ(M)$ is an RQ decomposition. 
 
 \noindent
 ii. 
An inverse-free update of $A_L(t)$ is found by solving 
\begin{equation}
    min_{\tilde{A}_L}|\tilde{A}_Lc(t+\delta t)-A_C(t+\delta t)|
     \label{eq:min A_L}
\end{equation}
where we have defined $A_C=A_Lc=cA_R$, with \\
$A_C(t+\delta t)=A_L(t)c(t)+\delta t \; d(A_Lc)/dt$ \\
and 
$c(t+\delta t)=c(t)+\delta t \; dc/dt$.
The time derivative of $A_L$ is obtained from Eq.(\ref{eq:TDVP}) and that of $c$ from
\begin{equation}
(\mathbb{I}-\sum^d_{\sigma=1}A^\sigma_L \otimes \bar{A}^\sigma_R)\frac{dc}{dt}=\sum^d_{\sigma=1}\frac{A^\sigma_L}{dt}cA^{\sigma \dagger}_R.
 \label{eq:c time evo}
\end{equation}
Eq.(\ref{eq:c time evo}) follows from writing $\frac{dc}{dt}=\frac{d}{dt}(\sum^d_{\sigma=1}A^\sigma_L c \bar{A}^\sigma_R)$ and using the right gauge fixing condition on $A_R$ to impose 
$\sum^d_{\sigma = 1} A_{ij}^{\sigma } c_{jk} dA^{\sigma \dagger}_{R,kl}=0$.
\noindent
iv. 
Eq.(\ref{eq:min A_L}) can be solved performing qr decompositions on $c(t+\delta t)$ and $A_C(t+\delta t)_{(\sigma i),j}$, $c(t+\delta t)=qr$ and $A_C(t+\delta t)_{(\sigma i),j}=QR$. We find $r=R$ so $A_L(t+\delta t)_{(\sigma i),j)}=Qq^\dagger$ and $A_L(t+\delta t)^\sigma_{ij}$ can be found by reshaping this matrix.

\subsection{Thermofield Double} 
The main modifications that we require to the standard MPS machinery is in its application to the thermofield double and its time evolution. 
We parametrize the thermofield double state $\ket{\psi \! \! \! \psi(A\! \! \! A)}$  by an expanded matrix product state $A \! \! \! A^{\sigma \delta}_{IJ}$ with a doubled physical index representing the two copies of the system. The thermofield double state is evolved using the expanded Hamiltonian, ${\cal H} \! \! \! \! {\cal H} = {\cal H} \otimes {\bm 1}+{\bm 1} \otimes {\cal H} $. The time-dependent variational principle
Eq.(\ref{eq:TDVPstate}) is modified accordingly with $A \rightarrow A \! \! \! A$, ${\cal H} \rightarrow {\cal H} \! \! \! \! {\cal H} $ and $\{i,j,\sigma\} \rightarrow \{ I,J,\sigma, \delta \}$. 

The thermofield double is evidently symmetrical between the two copies of the physical space; observations made on either copy will yield the same result. However, this is not necessarily reflected in an explicit symmetry of the tensor $A \! \! \! A^{\sigma \delta}_{I,J}$.  We therefore impose the symmetry $A \! \! \! A^{\sigma \delta}_{I,J}= A \! \! \! A^{\delta \sigma}_{\tilde I, \tilde J}$ on our state, where $I\equiv i\otimes i'$ and $ \tilde I \equiv i' \otimes i$, explicitly using an additional tangent space gauge fixing. This prevents a key source of errors: thermofield doubles form a subspace of the doubled Hilbert space. Strictly the dynamics under  ${\cal H} \! \! \! {\cal H}$ remains in this subspace, but numerical errors can take the dynamics away from it. 

For calculating our tangent state we find it more convenient to work in the a slightly different gauge in which the symmetry condition is   $A \! \! \! A^{\sigma \delta}_{I,J}=MA \! \! \! A^{\delta \sigma}_{\tilde I, \tilde J}M$, where  
\begin{equation}
M = \begin{pmatrix} \mathbb{I}_{\frac{D}{2}(D+1)} & 0 \\
0 & -\mathbb{I}_{\frac{D}{2}(D-1)},
\end{pmatrix}
\end{equation}

The tangent gauge fixing is then achieved as follows: We first calculate  $V \! \! \! \! V^{\sigma \delta}_{L,(IJ)}$ using the method described in section \ref{App:TDVP}. Symmetric ($\frac{1}{2}V \! \! \! \! V^{\sigma \delta}_{L,(I,J)}+M\frac{1}{2}V \! \! \! \! V^{\delta \sigma}_{L,( I  J)}$) and antisymmetric ($\frac{1}{2}V \! \! \! \! V^{\sigma \delta}_{L,(IJ)}-M\frac{1}{2}V \! \! \! \! V^{\delta \sigma}_{L,(I, J)}$) parts of $V \!\!\! \!V$ contribute separately to $d A \!\!\!A$ with corresponding symmetric and antisymmetric parts of the matrices $X$. The symmetrised  and anti-symmetrised spaces are each smaller than the doubled space. A complete orthonormal basis for $V \! \! \! \! V$ keeping the first  and a complete orthogonal basis for each is obtained by keeping the 
first $(d^2-1)D(D+1)/2-D$ or  $(d^2-1)D(D-1)/2+D$ (where $D\!\!\!\!D=D^2$)
columns of the Q from a QR decomposition of the symmetrised or anti-symmetrised $V \! \! \! \! V$ respectively. Full details of the implementation of this algorithm will be communicated elsewhere. This constraint also requires the modification of  step iv. in the inverse-free algorithm. $A \! \! \! A_L(t+\delta t)$ is calculated using QR decompositions on $c \! \! \! c(t+\delta t)$ and $A \! \! \! A_C(t+\delta t)$ but the symmetry constraint requires Q to be modified. A new Q is obtained by performing a QR decomposition on the symmetrised
$\frac{1}{2}Q_{((\sigma \delta) I),((\sigma \delta)' I')}+ \frac{1}{2}MQ_{((\delta \sigma) \tilde I),((\delta\sigma)' \tilde I')}M^\dagger$ and keeping the first $D\!\!\!\!D$ columns.


\section{Extraction of Lyapunov Spectrum}
\label{app:lTDVP}

In Appendix A we discussed how to calculate the Lyapunov spectrum of a trajectory in a dynamical system using vectors in its tangent space, in Appendix B we explained how time evolution of a quantum state can be determined using the time dependent variational principle, we will now explain how to extract the Lyapunov spectrum of a quantum system using these methods. As described in Section \ref{Sec:TDVP}, we are interested in the evolution of the difference of two trajectories, {\it i.e.} the tangent vectors to the variational manifold whose equation of motion is given by
 linearizing Eq.(\ref{eq:TDVP}) using the parametrization in terms of $X$ given by Eq.(\ref{eq:tangent space parameterization});
\begin{eqnarray}
d \dot X_{a}(t)
&=&
 i \langle \partial_{\bar{X}_{a}}  \partial_{\bar{X}_{b}}  \psi | \hat {\cal H} | \psi \rangle dX_{b}(t)
\nonumber\\
& &+
 i \langle \partial_{\bar{X}_{a}} \psi | \hat {\cal H} | \partial_{X_{b}}\psi \rangle d \bar X_{b}(t).
\label{eq:lTDVP C}
\end{eqnarray}
Our notation indicates a reshaping of the $(d^2-1)D \times D$ matrix $X$ into a complex $(d^2-1)D^2$ vector. In the case of thermofield double states, $X$ is complex $(d(d+1)/2-1)D^2$-dimensional vector. 

The right hand side of Eq.(\ref{eq:lTDVP C}) contains two parts: $F_{1}=\langle \partial_{\bar{X}_a} \psi | \hat {\cal H} | \partial_{X_b}\psi \rangle$ is manifestly Hermitian  and generates unitary rotations of the tangent vectors.  $F_{2}=\langle \partial_{\bar{X}_a}  \partial_{\bar{X}_b}  \psi | \hat {\cal H} | \psi \rangle$  is not Hermitian. Instead it is a symmetric matrix $F_{2}=F^{T}_{2}$ and is responsible for the non-unitary evolution of tangent vectors. 

The Lyapunov spectrum is calculated by measuring the extent to which a tangent vector $dX(t)$ has changed in magnitude between a time $t$ and $t+\delta t$. Eq.(\ref{eq:lTDVP C}) describes how the components $dX_{a}$  transform but doesn't account for the transformation of the tangent space basis. Taking in to account the transformation of the tangent space basis, we find
\begin{eqnarray*}
\frac{d (dX_a  \langle \partial_{X_a} \psi |)}{dt} 
&=& 
d \dot X_a \langle \partial_{X_a} \psi | + d X_a\frac{d \langle \partial_{X_a} \psi|}{dt}
\\
& =&
[d \dot X_a  - \bar{\Gamma}_{d,ab} dX_d \dot X_b]    \langle \partial_{X_a} \psi |
\end{eqnarray*}
where $\dot X^b$ is the time derivative of the coordinate along the trajectory and $\bar{\Gamma}_{a,bc}=\langle   \partial_{\bar{X}_b}  \partial_{\bar{X}_c}  \psi  | \partial_{X_a} \psi \rangle$ is the Christoffel symbol, which allows us to write 
$d \langle \partial_{\bar{X}_a} \psi|/dt=\bar{\Gamma}^{c}_{ab} \dot X_c \langle \partial_{[bar{X}_b} \psi|$. Note that in the parametrization of Eq.(\ref{eq:tangent space parameterization}), the metric is simply a delta function so that the difference between covariant and contravariant tensors is trivial.

Parallel transportation amounts to a modification to Eq.(\ref{eq:lTDVP C}) with $F_2$ being transformed to
\begin{equation}
F_2 \rightarrow \tilde{F}_2  = \langle \partial_{X_a}  \partial_{X_b}  \psi | \hat {\cal H} | \psi \rangle  - \bar{\Gamma}^c_{ab} \dot X_c 
\end{equation}
With this modification we can calculate the Lyapunov spectrum. We will separate the real and imaginary components of $dX=dX^R+idX^I$, $F_{1}=F^R_{1}+i F^I_{1}$ and $\tilde{F}_{2}=\tilde{F}^R_{2}+i \tilde{F}^I_{2}$. The real vector space is $2(d-1)D^2$ dimensional for matrix product states and  $(d(d+1)/2-2)D^2$ for thermofield double states.  Eq.(\ref{eq:lTDVP C}) can be rewritten as:
\begin{equation}
\begin{pmatrix} 
d\dot X^R \\
d\dot X^I
\end{pmatrix}
= \begin{pmatrix} 
F^I_{1}+\tilde{F}^I_{2}  & F^R_{1}-\tilde{F}^R_{2} \\
-F^R_{1}-\tilde{F}^R_{2}  & F^I_{1}-\tilde{F}^I_{2}
\end{pmatrix} 
\begin{pmatrix} 
dX^R \\
d X^I
\end{pmatrix}
\label{eq:lTDVP real}
\end{equation} 
If $\tilde{F}_2=0$ the Hermitian property of $F_1$ would result in a totally antisymmetric matrix in Eq.(\ref{eq:lTDVP real}), generating purely orthogonal rotations on the tangent vectors. $\tilde{F}_2$ is responsible for the changing magnitude of a tangent vector upon moving along a trajectory, and therefore for the generation of a non-zero Lyapunov spectrum. One important example in which $\tilde{F}_2=0$ is local Hamiltonians $H=\sum_i h_i$. In this case, the parallel transport term cancels with $F_2$, guaranteeing that the Lyapunov spectrum is zero for every possible state.

Having accounted for these details, the Lyapunov spectrum of the system can be calculated using Eq.(\ref{eq:lTDVP real}) and the methods in Appendix A.  

\end{document}